\documentclass[twocolumn,showpacs,prd,floatfix,aps,nofootinbib]{revtex4}
\usepackage{latexsym}
\begin{document}
\title{Restoration of the residue factorizability in the bound-state pole by instanton-antiinstanton configurations}
\author{Tomasz Rado\.zycki}
\affiliation{Department of Mathematical Methods in
Physics, Warsaw University, Ho\.za 74, 00-682 Warsaw,
Poland}
\email{torado@fuw.edu.pl} 

\begin{abstract}
The instanton-antiinstnton contributions to the $q\overline{q}$ bound state pole in the four-point Green function in the Schwinger Model are calculated. It is shown that these configurations, thanks to the cancellation of all unwanted terms, are responsible for the restoration of the perfect factorizability of the residue.
\end{abstract}
\maketitle

In gauge field theories the interplay between gauge group and space-time dimension may lead to the appearance of nontrivial topological effects, resulting from homotopy properties. The most important example of such a theory is Quantum Chromodynamics, where specific (instanton) gluon-field configurations play a significant role in the formation of the intricate vacuum and are related to various effects like chiral symmetry breaking and nonzero value of the quark condensate, provide solution of the $U(1)$ problem or influence quark interactions and meson or baryon correlators, to enumerate only few examples~\cite{huang,diak,mh,cnz,ss}. To clarify these issues much work has been devoted to the study of instantonic effects in model theories and particularly in the Schwinger Model~\cite{js,cadam1,smil,gattr,maie,rot,gmc,trinst}, which is highly nontrivial and --- due to its similarity to QCD in many aspects --- of real physical importance~\cite{kost}. The significance of this model for the investigations in hadronic physics should be emphasized especially as it allows exact and analytic results for a variety of interesting quantities. 

In our previous paper~\cite{trsing} we concentrated on the Bethe-Salpeter wave-function for the $q\overline{q}$ bound state, picked out from the polar term of the two-fermion Green function. This Green function was found exactly both with and without instanton contributions~\cite{lsb,trjmn,trinst}. The incorporation of higher instanton sectors (i.e. for instanton number $k=\pm 1,\pm 2$) turned out not only to modify the form of the Bethe-Salpeter function, which acquired additional terms, but also to destroy the factorization property of the residue in the bound state pole. This factorization was still maintained on the level of the $S$ matrix, but proved to be spoilt in the Green function. The other instantonic effect was the appearance of nonpolar, branch-point singularities, for $P^2=\mu^2$, where $\mu$ is the invariant mass of the bound state. The goal of this Brief Report is to show that the obtained nonfactorization of the bound-state pole residue may be cured, if one includes into consideration instanton-antiinstanton ($IA$) configurations.

The calculation of the $IA$ effects is more challenging than ordinary instanton calculus. Such a configuration formally bears $k=0$ topological number and there are only approximate fermionic zero modes of the Dirac operator, contrary to the latter case, where Atiyah-Singer index theorem guarantees the existence of true zero modes~\cite{as}. However, when the separation between instanton and antiinstanton becomes very large, these {\em quasi} zero modes approach real ones. Happily this is just the situation, we are interested in. Picking up the polar term out of the coordinate-space two-fermion Green function requires Fourier integration over infinite space, and the leading contribution is determined by the behaviour at space-time infinity. One can then make instanton-antiinstanton separation arbitrarily large.

Following~\cite{cadam1,trinst} we substitute into the path integral in the generating functional
\begin{equation}
Z[\eta,\overline{\eta},J]=\int{\cal D}\Psi{\cal D}\overline{\Psi}{\cal 
D}Ae^{i\int d^2x \left[{\cal 
L}+\overline{\eta}\Psi+\overline{\Psi}\eta+J^{\mu}A_{\mu}\right]}\; ,
\label{eq:genfunc}
\end{equation}
the following form of the gauge potential
\begin{equation}
A^\mu(x)=A^{(0)\mu}(x)+\varepsilon^{\mu\nu}\partial_{\nu}b(x)\; ,
\label{eq:genpot}
\end{equation}
with $\varepsilon^{\mu\nu}$ being the antisymmetric symbol.
The integration, which will be led in euclidean space, is now performed over $A^{(0)\mu}$ and is restricted to the instanton sector $k=0$, which requires the topologically trivial behaviour of the gauge field at infinity. The substitution~(\ref{eq:genpot}) is a simple shift and does not influence the integration measure. The background function $b$ is chosen below for our convenience. ${\cal L}$ is the standard Schwinger Model Lagrangian
\begin{equation} 
{\cal L}(x)=\overline{\Psi}(x)\left(i\gamma^{\mu}\partial_{\mu} -
g\gamma^{\mu}A_{\mu}(x)\right)\Psi (x)-
\frac{1}{4}F^{\mu\nu}(x)F_{\mu\nu}(x)\; . 
\label{eq:lagr} 
\end{equation} 
with a gauge fixing term, if needed.\footnote{Green functions are, naturally, gauge dependent. The following formulae will be given in the Landau gauge.} Since we are interested in the $IA$ configuration, $b$ will be taken as a simple sum of the fields corresponding to pure instanton and pure antiinstanton at shifted positions~\cite{lev,guida}.  The function $b$ has, therefore, the following (euclidean) form
\begin{eqnarray}
b_{(+-)}(x)&=&\frac{i}{2g}\ln\left(\frac{(x+R/2)^2+\rho^2}{\rho^2}\right)\nonumber\\
&-&\frac{i}{2g}\ln\left(\frac{(x-R/2)^2+\rho^2}{\rho^2}\right)
\; ,
\label{eq:b}
\end{eqnarray}
where $(+-)$ refers to the $IA$ configuration. For $AI$ one, we will have $b_{(-+)}(x)=-b_{(+-)}(x)$. The separation of $I$ and $A$ is set equal to $R$ and we will be interested in the limit $R\rightarrow \infty$. 

The functional integration over $A^{(0)\mu}$ in (\ref{eq:genfunc}) is simple. First we have to gauge away the term $g\overline{\Psi}\not\!\!\! A^{(0)}\Psi$ from the Lagrangian by the appropriate redefinition of fermion fields (which leads to the appearance of the gauge-boson mass term with $\mu^2=g^2/\pi$). Next $A^{(0)\mu}$ appearing in source terms (because of the above redefinition) is replaced with the functional derivative over external current $J_\mu$ and may be driven out from under the integral. The remaining integral becomes then Gaussian and easy to be taken. The details of this calculation are given in~\cite{trinst} and are not dependent on the specific choice of the function $b$, hence there is no need to repeat them here. 

The next step is to perform fermion integrals. They are more complicated but, fortunately, the majority of the work has already been done in~\cite{trinst}. We limit ourselves to the four-point Green function, which is the fourth-order coefficient of the expansion of~(\ref{eq:genfunc}) in powers of fermionic sources $\eta$ and $\overline{\eta}$ (i.e. it is $G$ from the term $\overline{\eta}\,\overline{\eta}\,G\,\eta\,\eta$). This four-point function was found exactly with all the instantonic corrections ($k=0,\pm 1,\pm 2$), as well as the form of the $q\overline{q}$ bound state pole and the appropriate formulae may be found in~\cite{trinst,trsing}. What we would like to concentrate on below, is the sector $k=0$ with the background configuration~(\ref{eq:b}), when $R$ approaches infinity. This calculation is somewhat similar to that for $k=\pm 2$ in the sense that we have again two ({\em quasi}) zero modes, which (both) have to appear in the Green function due to the rules of Grassman integration (cf. formula~(51) of~\cite{trinst}). We are only concerned with the leading asymptotic term, so these two {\em quasi} zero modes may be taken as that of the pure instanton centered on $-R/2$ and that of the pure antiinstanton centered on $R/2$~\cite{cadam1,maie,jaye}. Obviously, for finite $R$ they are not exact.
\begin{eqnarray}
\chi_1(x)&=&\frac{1}{\sqrt{2\pi}}\left(\frac{1}{(x+R/2)^2+\rho^2} 
\right)^{1/2}\left(\begin{array}{c}0 \\ 1\end{array}\right)\; ,\label{eq:zerop}\\
\chi_2(x)&=&\frac{1}{\sqrt{2\pi}}\left(\frac{1}{(x-R/2)^2+\rho^2} 
\right)^{1/2}\left(\begin{array}{c}1 \\ 0\end{array}\right)\;,\label{eq:zerom}
\end{eqnarray}
They are eigenfunctions (with eigenvalues $\pm 1$) of the chirality operator $\gamma^5$ (our  conventions as to the $\gamma$ matrices are given in~\cite{trinst}). 

Now our previous calculation can be applied with only obvious modifications and with the additional difference that our formulae will become precise only in the limit $R\rightarrow \infty$. Referring the Reader to~\cite{trinst} and without going into details, we can write down the formula for the $IA$ contribution to the Green function in question
\begin{widetext}
\begin{eqnarray}
G^{IA}_{ab;cd}(x_1,x_2;x_3,x_4)=&-&\frac{1}{\lambda_1\lambda_2}\bigg[\chi_{1a}^{ }(x_1)\chi_{1c}^+(x_3)\chi_{2b}^{ }(x_2)\chi_{2d}^+(x_4)e^{-ig(b_{(+-)}(x_1)+b_{(+-)}(x_3)-b_{(+-)}(x_2)-b_{(+-)}(x_4))}\nonumber\\&+&
\chi_{2a}^{ }(x_1)\chi_{2c}^+(x_3)\chi_{1b}^{ }(x_2)\chi_{1d}^+(x_4)e^{ig(b_{(+-)}(x_1)+b_{(+-)}(x_3)-b_{(+-)}(x_2)-b_{(+-)}(x_4))}\bigg]\label{eq:gia}\\
&\times&e^{ig^2\left[\beta(x_1-x_3)+\beta(x_2-x_4)-\beta(x_1-x_4)-\beta(x_1-x_2)-\beta(x_2-x_3) - 
\beta(x_3-x_4)\right]}- \left\{\begin{array}{ccc} c&\leftrightarrow&d\\ x_3 &\leftrightarrow & x_4\end{array}\right\}\nonumber\; .
\end{eqnarray}
\end{widetext}
Here $a,b,c,d$ are spinor indices, the function $\beta$ was defined in~\cite{trinst} and $\lambda_{1,2}$ are eigenvalues for two {\em quasi} zero modes of the Dirac operator given below  in~(\ref{eq:dia}) and appear in denominator due to the normalization of the vacuum-vacuum transition amplitude to unity in the $k=0$ topological sector (other sectors do not contribute to this amplitude if fermions are massless, since the tunneling between various topological vacua is suppressed by zero eigenvalues). We would like to stress here again that $\chi^{ }_{1,2}$ {\em are not} eigenvectors with eigenvalues $\lambda_{1,2}$, but approach them when $R$ tends to infinity. Neither {\em quasi} zero modes nor their eigenvalues $\lambda_{1,2}$ have to be known exactly, but it is sufficient to know their asymptotic behaviour (naturally $\lambda_{1,2}$ tend to zero).

Now we have to estimate these eigenvalues for large $R$. In this limit the true {\em quasi} zero modes can be chosen to be arbitrarily close to $\chi^{ }_{1,2}$. Therefore, following~\cite{sv,ss,kov} we assume, that the subspace in question is spanned just by $\chi^{ }_1$ and $\chi^{ }_2$. The euclidean Dirac operator in the $IA$ background has the form
\begin{widetext}
\begin{eqnarray}
D^{IA}&=&\left(\begin{array}{cc} 0&i\partial_2-\partial_1\\ i\partial_2+\partial_1&0\end{array}\right)-\frac{1}{(x+R/2)^2+\rho^2}\left(\begin{array}{cc} 0&x_1+R_1/2-i(x_2+R_2/2)\\ x_1+R_1/2+i(x_2+R_2/2)&0\end{array}\right)\nonumber\\
&+&\frac{1}{(x-R/2)^2+\rho^2}\left(\begin{array}{cc} 0&x_1-R_1/2-i(x_2-R_2/2)\\ x_1-R_1/2+i(x_2-R_2/2)&0\end{array}\right)\; .
\label{eq:dia}
\end{eqnarray}
\end{widetext}
In the vector space with the scalar product defined as $<\!f|g\!>=\int d^2xf^+(x)g(x)$, it is a hermitian operator with only off-diagonal matrix elements
\begin{equation}
\left(\begin{array}{cc} 0& D^{IA}_{1,2}\\ D^{IA}_{2,1} & 0\end{array}\right)
\end{equation}
and eigenvalues equal to $\pm\sqrt{D^{IA}_{1,2}D^{IA}_{2,1}}$, where indices 1 and 2 refer to the functions~(\ref{eq:zerop}) and~(\ref{eq:zerom}). The diagonal elements disappear since~(\ref{eq:dia}) inverts the chirality and $\chi^{ }_{1,2}$ are chirality eigenvectors. 

One can easily verify, that $D^{IA}_{1,2}=\overline{D^{IA}_{2,1}}$, which means that the eigenvalues are real, as necessary for the hermitian operator. These eigenvalues should be substituted for $\lambda_{1,2}$ in~(\ref{eq:gia}), but what we need, is only their asymptotic form. For instance we have
\begin{eqnarray}
D^{IA}_{2,1}&=&\int d^2x\chi^+_2(x)D^{IA}\chi_1(x)\label{eq:dia21}\\
&=&\frac{1}{2\pi}\int d^2x\frac{x_1-ix_2}{(x^2+\rho^2)^{3/2}((x+R)^2+\rho^2)^{1/2}}\; ,\nonumber
\end{eqnarray}
where the integration variable has been shifted by $R/2$. This integral may be performed using the method of Feynman parameters, usually applied in the calculation of Feynman diagrams, and we obtain 
\begin{equation}
D^{IA}_{2,1}=-\frac{1}{\pi}\int_0^1d\alpha\frac{\sqrt{\alpha(1-\alpha)}(R_1-iR_2)}{R^2\alpha(1-\alpha)+\rho^2}\sim -\frac{1}{R_1+iR_2}\; ,
\end{equation}
which leads to the asymptotic form for the product of eigenvalues $\lambda_1\cdot\lambda_2\sim \frac{1}{R^2}$.

Now we can come back to~(\ref{eq:gia}). To find the polar contribution, we introduce new variables, as it was done in~\cite{trsing}
\begin{eqnarray}
X=\frac{1}{2}(x_1+x_3)\; ,&&\;\;\;\; x=x_1-x_3\; ,\nonumber\\*
Y=\frac{1}{2}(x_2+x_4)\; ,&&\;\;\;\; y=x_2-x_4\; .\label{eq:coord}
\end{eqnarray}
The translational invariance should manifest itself through the dependence of $G$ on $x_i-x_j$'s only or, in new variables, on $Z=Y-X$, $x$ and $y$. This is not visible in~(\ref{eq:gia}) since, for finite $R$, our formulae are only approximate ones. Because we are interested in the $t$-channel singularity, the pole corresponding to the bound state should be found in the complex plane of $P^2$, where $P$ is the two-momentum canonically conjugated to $Z$. The $Z$ dependence of the expression is then crucial.

Let us begin with the first term in the square brackets in~(\ref{eq:gia}). Its matrical structure, according to~(\ref{eq:zerop}) and~(\ref{eq:zerom}), has the form $(\openone-\gamma^5)_{ac} (\openone+\gamma^5)_{bd}$. To reveal the dependence on $Z$ we first have to simplify the expression exploiting~(\ref{eq:b}),~(\ref{eq:zerop}) and (\ref{eq:zerom}), which leads to the explicit cancellation of several factors coming from  $b_{(+-)}$ and $\chi^{ }_{1,2}$. Next the $Z$-dependent function should be isolated of the last exponent (that containing $\beta$ functions). Up to the opposite sign, this was actually done in~\cite{trsing} (cf. formulae~(26),~(27) and following ones). The $Z$ dependence is there twofold: firstly in the factor (with altered sign)
\begin{equation}
\exp[\ln(-\mu^2 Z^2/4)]=-\frac{\mu^2 Z^2}{4}\; ,
\label{eq:pref}
\end{equation}
and secondly in $e^{-ipZ}$, which (for the polar term) simply shifts the Fourier variable $P$. The rest of the expression (i.e. the pole itself, although, before taking the Fourier transform, in the  variable $p$ ) may be derived from the formula~(30). 

Gathering all essential factors we see, that the function (still written in euclidean space) to be Fourier transformed is

\begin{equation}
\frac{R^2Z^2}{\sigma^{+}_x\sigma^{-}_x\sigma^{+}_y\sigma^{-}_y}\; ,
\label{eq:factor}
\end{equation}
where $\sigma^{\pm}_x=((X\pm x/2-R/2)^2+\rho^2)^{1/2}$ and $\sigma^{\pm}_y=((Y\pm y/2+R/2)^2+\rho^2)^{1/2}$.

For the limit of infinite $R$ the relative variables $x$ and $y$ become unessential and may be omitted. $X$ and $Y$, in turn, can be rewritten as 
\begin{eqnarray}
X&=&(X+Y)/2-Z/2\nonumber\\
Y&=&(X+Y)/2+Z/2\; .\nonumber
\end{eqnarray}
The polar term is connected only with the infinite $Z$ integration, so $(X+Y)/2$ may again be disregarded as $R\rightarrow \infty$. This leads to the simplified expression 
\begin{equation}
\frac{16 R^2Z^2}{(Z+R)^2+4\rho^2)^2}\; .
\label{eq:facsimp}
\end{equation}
Now we are in a position to perform the Fourier integration over $Z$, together with the limit over $R$. They should be done in such a way that $Z$ and $R$ are always close to each other ($(Z-R)^2<<R^2$), so one can substitute $Z=R+w$ in~(\ref{eq:facsimp}), take the asymptotic term as $R$ goes to infinity, and finally perform the Fourier integral over new variable $w$. The Fourier factor under the integral is $e^{i(P-p)Z}$. Analyzing~(\ref{eq:facsimp}) one sees that asymptotically the integrand function becomes unity and, therefore, the leading term of the Fourier transform is simply $(2\pi)^2\delta^{(2)}(P-p)$. 

In an analogous way one can easily verify that the second term in the square brackets in~(\ref{eq:gia}) does not contribute to the pole. The term obtained by the anti-symmetrization, in turn, contributes only to the pole in the $u$-channel. 

We can now gather the whole expression for the $IA$ configuration
\begin{eqnarray}
iG^{IA}_{b.s.}(P;x,y)=-\frac{\mu^2}{8\pi}\frac{\cos(Px/2)\cos(Py/2)}{P^2-\mu^2+i\epsilon}&&\label{eq:giabs}\\
\times\, e^{2\gamma_E}e^{ig^2(\beta(x)+\beta(y))}(\openone-\gamma^5)\otimes (\openone+\gamma^5)&&,\nonumber
\end{eqnarray}
where $b.s.$ stands for `bound state'. The coefficient $\mu^2$ comes from~(\ref{eq:pref}) and cosine functions, together with the pole, from the expansion of the last exponent in~(\ref{eq:gia}) --- except of the first two $\beta$'s, which do not depend on $Z$ --- similarly as it was done in~\cite{trsing}.

Now we should consider the $AI$ configuration. This contribution may be found in an identical way and there is no need to repeat all steps again. It simply corresponds to the substitution $R\rightarrow -R$, so now the second term in brackets of~(\ref{eq:gia}) has the appropriate limit and contributes to the pole. The straightforward calculation leads to the similar result as above only with the modification in the $\gamma$ matrices structure, which actually might be predicted. The two modes~(\ref{eq:zerop}) and~(\ref{eq:zerom}) exchange their roles, so the matrical structure is changed into $(\openone+\gamma^5)\otimes (\openone-\gamma^5)$ and the rest of the expression~(\ref{eq:giabs}) remains unaltered. The whole bound state pole contribution to the Green function is then
\begin{eqnarray}
&&\!\!\!\!\!\!\!\! iG^{IA+AI}_{b.s.}(P;x,y)\label{eq:giatot}\\
&&=-\frac{\mu^2}{8\pi}\frac{\cos(Px/2)\cos(Py/2)}{P^2-\mu^2+i\epsilon} e^{2\gamma_E}e^{ig^2(\beta(x)+\beta(y))}\nonumber\\
&&\times\left[(\openone-\gamma^5)\otimes (\openone+\gamma^5)+(\openone-\gamma^5)\otimes (\openone+\gamma^5)\right]\; .\nonumber
\end{eqnarray}
Confronting the obtained result with that of our previous work~\cite{trsing} we observe that~(\ref{eq:giatot}) exactly cancels all the unwanted terms in the formula~(32) and restores the full factorizability already on the level of the Green function. (The absence of $\cos\frac{Px}{2}$ in the formula~(33) is only a typing omission.) The only apparent difference is in $\theta$ dependence via the two factors $e^{-i\theta\gamma^5}$ arising in instanton sector $k=\pm 2$ and certainly absent for the $IA$ and $AI$ configurations, which belong to $k=0$ sector. Actually, since both factors are accompanied by $(\openone\pm\gamma^5)$, and because $\gamma^5(\openone\pm\gamma^5)=\pm(\openone\pm\gamma^5)$, they simply reduce to $e^{\mp i\theta}$. In consequence, appearing always in opposite pairs, they cancel each other. Anyway, in the massless theory, $\theta$ does not play any role and may be gauged away or simply set to 0. The above calculation reveals again the complexity of the field theory vacuum  with underlying nontrivial topological structure.

\end{document}